\documentclass[conference]{IEEEtran}
\ifCLASSINFOpdf
\else
   \usepackage[dvips]{graphicx}
  \graphicspath{{G:\Documents and Settings\Behrman\My Documents\papers\singapore}}
  \DeclareGraphicsExtensions{.eps}
\fi
\begin{document}
%
\title{A quantum neural network computes its own relative phase }


\author{\IEEEauthorblockN{Elizabeth C. Behrman}
\IEEEauthorblockA{Department of Mathematics, Statistics, and Physics\\
Wichita State University\\
Wichita, Kansas 67260--0033\\
email: elizabeth.behrman@wichita.edu}
\and
\IEEEauthorblockN{James E. Steck}
\IEEEauthorblockA{Department of Aerospace Engineering\\
Wichita State University\\
Wichita, Kansas 67260--0044\\
email: james.steck@wichita.edu}}


%


\maketitle

\begin{abstract}
Complete characterization of the state of a quantum system made up of subsystems requires determination of relative phase, 
because of interference effects between the subsystems. For a system of qubits used as a quantum computer this is especially vital, because 
the entanglement, which is the basis for the quantum advantage in computing, depends intricately on phase. 
We present here a first step towards that determination, in which we use a two-qubit quantum system as a quantum neural network, 
which is trained to compute and output its own relative phase.
\end{abstract}


%
\IEEEpeerreviewmaketitle

\section{Introduction}
 Entanglement is the root of the power of quantum computers\cite{genentref}; thus, the production and measurement of entanglement are essential if we are ever to be successful in making full use of the potential of quantum computing. This turns out to be a very hard problem.  

In previous work, we have proposed a method to find an entanglement witness for a general, unknown, quantum input state, using dynamic  learning to find parameters for the quantum system that make it calculate its own entanglement. We called this a quantum neural network (QNN) \cite{eb08}. The basic idea is that contained in the system itself is the information about its entanglement: If we find, through learning, an appropriate set of parameters for the system, then it can extract the entanglement of its initial state as an output measure of the state at some final time. We imagine that our quantum system evolves under some Hamiltonian containing adjustable parameters; we find that set of parameters such that our designated output function (the qubit-qubit correlation function) is mapped onto the correct values for the entanglement of the initial state. Our entanglement witness gave good results for large classes of input states, including both pure and mixed states. Unlike the case with any other witness (see, {\it e.g.}, \cite{toth}), the input state did not need to be ``close'' to any particular state. We have also \cite{nabic,eb12} extended our work to the 3-, 4-, and 5-qubit cases, and found that as the size of the system grows, the amount of additional training necessary diminishes; thus, our method may be very practical for use on large computational systems.

Figure 1 shows some representative results, in which we compare our entanglement witness to the entanglement of formation \cite{wootters} for 50,000 randomly generated states for the 2-qubit system. These are pure states with real coefficients on the usual (``charge'') basis. The agreement is excellent. Unfortunately these results do not carry over to the more general case of complex coefficients. See Figure 2, which shows a similar set but with complex coefficients.  Indeed, as we showed \cite{eb08}, it is impossible to find any single measurable which will not exhibit anomalous oscillation; all witnesses do so. But is there a way to get around this difficulty?

\begin{figure}[!t]   
\centering
\includegraphics[width=2.5in]{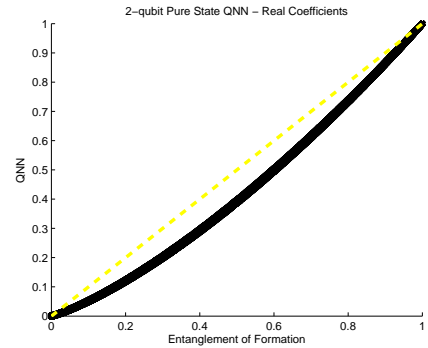}
\caption{QNN entanglement for 50,000 randomly generated pure states of the form $a_{00}|00> + a_{01}|01> + a_{10}|10> + a_{11}|11>$, where $a_{00}$, $a_{01}$, $a_{10}$,and $a_{11}$ are all real, as a function of the entanglement of formation. Points lying along the dashed yellow line are states for which the entanglement predicted by the QNN witness exactly matches the entanglement of formation. }
\label{realfig}
\end{figure}

\begin{figure}   
\centering
\includegraphics[width=2.5in]{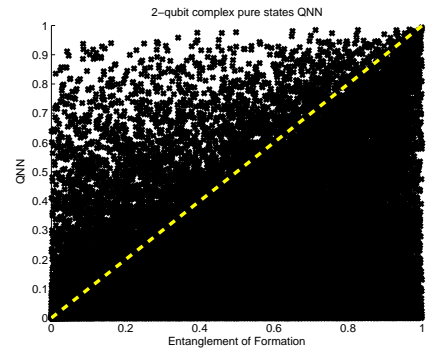}
\caption{As in Figure \ref{realfig}, but with complex coefficients. \label{cxcoeffig}}
\end{figure}

There are, of course, ways to determine more information about the state; if we know the entire density matrix we can, at least for the 2-qubit system, simply calculate the entanglement of formation (as we ourselves did to generate the comparison data for Figure \ref{realfig}.) For the 2-qubit system this may not be unreasonable. But for the eventual goal of a large computational system, this can become quite daunting, since the number of parameters necessary goes like $2^{2N}$, where $N$ is the number of qubits. Perhaps dynamic learning can allow us to find a shortcut. This paper is a first step in that direction.

If we knew or could determine the relative phases $\{\theta\}$ of the basis states, we could apply the (unitary) phase shift operator of $\{e^{-i \theta} \}$ to each relevant part of our input state. Since the coefficients would then be real, we could then perform our entanglement witness measurement and achieve results like those in Figure \ref{realfig}.

In 2005, Yang and Han \cite{han} found an algorithm for determining the relative phase between parts of the n-qubit Bell (or GHZ) state, $ \sqrt{p}|0...0> + e^{i \phi} \sqrt{1-p}|1...1>$.  They showed that performing a Hadmard transform on each qubit puts the system in a state in which the probability of finding an even number of qubits in the state $|1>$ is given by $p_{even} = \frac{1}{2} + \sqrt{p(1-p)} \cos{\phi}$. Given a large number of copies of the state, it is then possible to determine both $p$ and $\phi$.  Here we show that, with our QNN, we can extend this result, for the 2-qubit system,  in two ways. First, we show that we can also find the phase offset in an EPR state, $ a_{01}|01> + e^{i \theta}a_{10}|10>$. Second, we show that we can also find the phase offset for any of the partially entangled states consisting of an EPR or a Bell state with some contaminant:
\begin{eqnarray}
a_{00}|00> + a_{01}|01> + e^{i \phi}a_{11}|11> 
\\ \nonumber a_{00}|00> + a_{10}|101> + e^{i \phi}a_{11}|11>
\\ \nonumber a_{00}|00> + a_{01}|01> + e^{i \theta}a_{10}|10>
\\ \nonumber a_{01}|01> + e^{i \theta}a_{10}|10> + a_{11}|11>
\\ \nonumber e^{i \xi}a_{01}|01> + a_{10}|10>+ a_{11}|11> 
\\ \nonumber a_{00}|00> + e^{i \xi}a_{01}|01> + a_{10}|10>
\end{eqnarray}

\section{Dynamic learning: quantum neural network (QNN)\label{qnn}}

We consider 2-qubit system whose Hamiltonian is:
\begin{equation}
H = K_{A} \sigma_{xA} + K_{B} \sigma_{xB} + \varepsilon_{A} \sigma_{zA} + \varepsilon_{B} \sigma_{zB} + \zeta \sigma_{zA} \sigma_{zB} 
\label{hamiltonian}
\end{equation}
where $\{ \sigma \}$ are the Pauli operators corresponding to each of the two qubits, A and B, $K_{A}$ and $K_{B}$ are the tunneling amplitudes, $\varepsilon_{A}$ and $\varepsilon_{B}$  are the biases, and $\zeta$ the qubit-qubit coupling. The time evolution of the system is then given by the  Schr\"{o}dinger equation:
\begin{equation}
\frac{d \rho}{dt} = \frac{1}{i \hbar}[H, \rho] 
\label{schr}
\end{equation}
where $\rho$ is the density matrix and $H$ is the Hamiltonian. The parameters $\{ K,\varepsilon,\zeta \}$ control the time evolution of the system in the sense that, if one or more of them is changed, the way a given state will evolve in time will also change. This is the basis for using our quantum system as a neural network. The role of the ``weights'' of the network is played by the parameters of the Hamiltonian, $\{ K,\varepsilon,\zeta  \}$, all of which we take to be  experimentally adjustable as functions of time (see, {\it e.g.}, \cite{yamamoto}, for the case of SQuID charge qubits.) By adjusting the parameters using a neural  learning algorithm we can train the system to evolve in time to a set of chosen target outputs at the final time $t_{f}$, in response to a corresponding (one-to-one) set of given inputs. Because the time evolution is quantum mechanical (and, we assume, coherent), a quantum mechanical function, like an entanglement witness of the initial state, can be mapped to an observable of the system's final state, a measurement made at the final time $t_{f}$. The time evolution of the quantum system is calculated by integrating the Schr\"{o}dinger equation numerically in MATLAB Simulink, using ODE4 (Runge-Kutta), with a fixed integration step size of 0.05 ns \cite{matlab}. The system was initialized in each input state in the training set, in turn, then allowed to evolve for 190 ns. A measurement is then made at the final time; this is the ``output'' of the network. An error, $target-output$, is calculated, and the parameters are adjusted slightly to reduce the error. This is repeated for each $(input,target)$ pair multiple times until the calculation converges on parameters that work well for the entire training set.  Complete details, including a derivation of the quantum dynamic learning paradigm using backpropagation \cite{lecun} in time \cite{werbos}, are given in \cite{eb08}.

 We choose the usual ``charge basis '', in which each qubit's state is given as 0 or 1. 

 All of the parameters $\{ K,\varepsilon,\zeta \}$ were taken to be functions of time; in contrast to our earlier work \cite{eb08,nabic,eb12}, in which the parameters were taken to be piecewise constant in time,  we have, here, allowed the parameters to be continuous functions of time. For the backpropagation learning, the output error needs to be back-propagated backward through time \cite{werbos}, so the integration has to be carried out from the final time $t_{f}$ to $0$.  To implement this in MATLAB Simulink, a change of variable is made by letting $t' = t_{f}-t$, and running this simulation forward in $t'$ in Simulink.

\section{Training of the phase indicator}

In the charge basis, we can write a general pure state of the system at time $t=0$ as 
\begin{eqnarray}
|\Psi(0)>=a_{00}|00>+a_{01}e^{i\xi}|01>+a_{10}e^{i \theta}|10> \\ \nonumber +a_{11}e^{i \phi}|11>
\end{eqnarray}
where normalization requires that 
\begin{equation}
\sqrt{a_{00}^{2}+a_{01}^{2}+a_{10}^{2}+a_{11}^{2}}=1
\end{equation}
Since an \underline{overall} phase is physically meaningless we may take out any overall phase factor; that is, without loss of generality we may take the coefficient of the $|00>$ basis state to be real. We then write each of the other coefficients as its magnitude times a phase factor; thus, each $a_{nm}$ will be a real number, and the phase factor, if any, will be written in explicitly. As discussed above, the state of the system evolves under the Hamiltonian, Equation \ref{hamiltonian}, to another state, $|\Psi(t_{f})>$, at the final time $t_{f}$. At that final time we make a measurement. In the terminology of neural network learning: (1) the input to the neural network is the initial state $|\Psi(0)>$ at time $t=0$ of the quantum system; (2) the output of the neural network is a quantum measure made on the final state at the final time $t=t_{f}$ of the quantum system; and (3) the trainable weights of the neural network are the time histories of the adjustable parameters of the quantum system.  The network is trained on a set of training pairs, each of which consists of (input, correct output).  Each training pair is presented to the network, the output is calculated, the error computed, and the weights changed so as to decrease the error \cite{eb08}.  Each pass through the entire training set is called an epoch. 

As with all good science, we began with what was already known \cite{han}: namely, that it is possible to extract relative phase information from the Bell state, $|Bell>= a_{00}|00> + a_{11}e^{i \phi}|11>$. Because we are using a \underline{learning} process, it is important to see how much information we can get with as little input as possible. Thus, our original training set consisted of only 11 training pairs, using only equal amplitude Bell states $|Bell>= a_{00}|00> + a_{11}e^{i \phi}|11>$ with $a_{00}=a_{11}=\frac{1}{\sqrt{2}}$, where the phase angle $\phi$ varies from $-\pi/2$ to $\pi/2$ as $\phi = -\frac{\pi}{2} + \frac{(n-1)\pi}{10}$, for $n=1:11$. The network output is the absolute magnitude squared of the projection of the final state of the quantum system onto the state $|11>$, {\it i.e.}, the probability of the system's being found in the state $|11>$.  The correct, or target, output for these equal-amplitude EPR states is taken to be just $\cos^{2}(\phi/2)$. That is, the (input,output) pairs are 
\begin{eqnarray}
input =  |\Psi(0)>= \frac{1}{\sqrt{2}}(|00> + e^{i \phi}|11>) \\ \nonumber
output = |<11|\Psi(t_{f})>|^{2} \rightarrow target = \cos^{2}(\phi/2)
\end{eqnarray} 

\begin{figure}   
\centering
\includegraphics[width=2.5in]{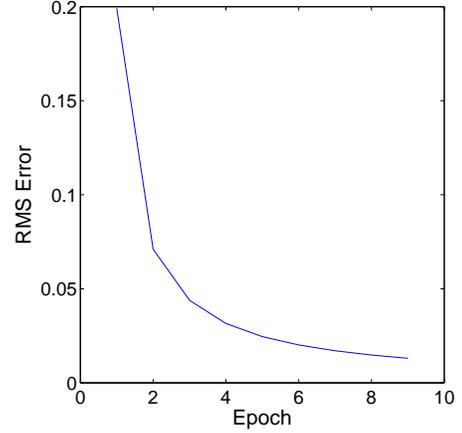}
\caption{RMS error per training pair vs. epoch (pass through the training set) for the $\phi$ phase offset indicator. The training set of 11 (input,output) pairs is given in the text. \label{trainBfig}}
\end{figure}

\begin{figure}  
\centering
\includegraphics[width=2.5in]{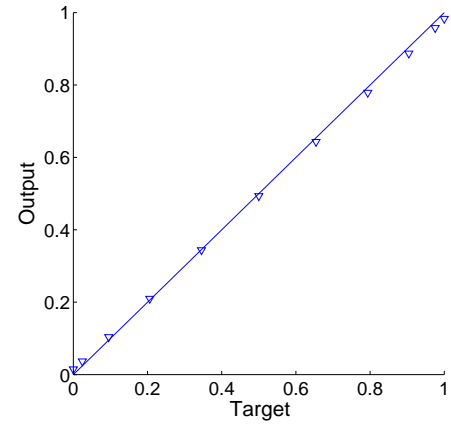}
\caption{Results for the training set for the $\theta$ phase offset indicator, showing deviation of the output, $|<11|\Psi(t_{f})>|^{2}$, from the target function $\cos^{2}(\phi/2)$. Average RMS error per pair is 0.0127. The line shows the goal (perfect agreement.) \label{trainresBfig}}
\end{figure}

\begin{figure}   
\centering
\includegraphics[width=2.5in]{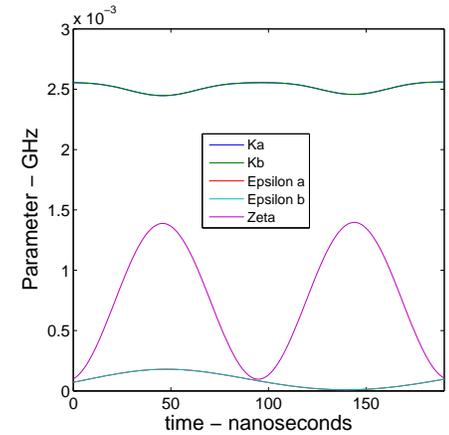}
\caption{The functions $K_{A}$, $K_{B}$, $\epsilon_{A}$, $\epsilon_{B}$, and $\zeta$, as functions of time, as trained for the phase offset $\theta$. $K_{A}$ and $K_{B}$ lie right on top of each other, as do $\epsilon_{A}$ and $\epsilon_{B}$. Each was started out (pre-training values) as constant functions:  $K_{A} = K_{B}= 2.5 \times 10^{-3} GHz$ , and $\epsilon_{A}=\epsilon_{B}=\zeta=10^{-4} GHz$. \label{paramBfig}}
\end{figure}

The network was trained for 10 epochs, on a total of 11 training pairs.  The average RMS error of all 11 training pairs after training is 0.0127.  A plot of RMS error vs epoch is shown in Figure \ref{trainBfig}. A plot of output vs target for the 11 training pairs is shown in Figure \ref{trainresBfig}.  A plot of the trained parameters as functions of time is shown in Figure \ref{paramBfig}. Each is a simple oscillatory function. Note that the trained tunneling amplitude functions $K_{A}$ and $K_{B}$ lie right on top of each other, as do $\epsilon_{A}$ and $\epsilon_{B}$, which is unsurprising given the symmetry of the training set.

\begin{figure}  
\centering
\includegraphics[width=2.5in]{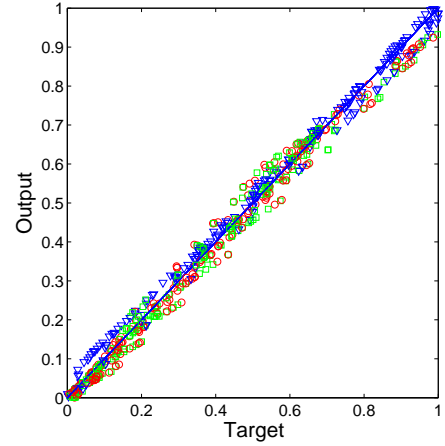}
\caption{Results for the testing set for the phase offset $\phi$, consisting of three types of states: unequal amplitudes Bell states $a_{00}|00> + e^{i\phi}a_{11}|11>$ (blue triangles), and two kinds of partially entangled states, $a_{00}|00> + a_{01}|01> + e^{i \phi}a_{11}|11>$ (green squares), and  $a_{00}|00> + a_{10}|101> + e^{i \phi}a_{11}|11>$  (red circles).  As in Figure \ref{trainresBfig}, we show the deviation of the output from the target functions (given in the text.) Average RMS error per pair is 0.0270. The line shows the goal (perfect agreement.) \label{test3fig}}
\end{figure}

To see if the network has generalized ({\it i.e.}, \underline{learned} as opposed to having simply curvefitted), we then tested (with no additional training) on a set of Bell states of random relative magnitude, that is, on states of the type $|Bell>= a_{00}|00> + a_{11}e^{i \phi}|11>$ with now randomly generated numbers for $a_{00}$ and $a_{11}$ (such that the state remained normalized, {\it i.e.}, $\sqrt{a_{00}^{2}+a_{11}^{2}} = 1$.) From \cite{han} we knew that it was unlikely that we would be able to train to the same simple target function $\cos^{2}(\phi/2)$ , and so it transpired; however, we found that a simple analogue, $2(\frac{1}{2} - a_{00}^2)^{2} a_{11}^{2} + 2a_{00}a_{11}\cos^{2}(\phi/2)$, did work quite well. Again, the measurable is the probability of the system's being found in the $|11>$ state at the final time, {\it i.e.},$|<11|\Psi(t_{f})>|^{2}$. Note that this target function reduces to the target function used for training, when $a_{00}=a_{11}=\frac{1}{\sqrt{2}}$, and maintains the necessary symmetry. These data are plotted in Figure \ref{test3fig} (blue triangles.) As can easily be seen in the figure, agreement is excellent. 

The ability of the system to map onto the target function depends on its being an entangled state \cite{han}; however, as long as we adjust the target function appropriately, full entanglement is, clearly, not necessary. Thus it ought also to be possible to find the phase offset for a partially entangled input state, {\it e.g.}, of the form $a_{00}|00> + a_{01}|01> + e^{i \phi}a_{11}|11>$. How do we do this? We consider a probability-weighted target function, equal to our earlier targets for the special case of the pure Bell state, but adjusting the relative function for the diminished entanglement. We are guided here by symmetry and by earlier analytic results \cite{han}, in which it was found that, while the relative phase was extractable, it was not easily separable from the amplitude information, and, in fact, had to be separately measured for (hence, the necessity for ``many copies'' of the original state.) Experimentation eventually gave us the following (relatively) simple functions. For the Bell state, $a_{00}|00> + a_{11}e^{i\phi}|11>$, the target function for the output $|<11|\Psi(t_{f})>|^{2}$ is:  
\begin{equation}
target_{Bell} =  2(\frac{1}{2} - a_{00}^2)^{2} a_{11}^{2} + 2a_{00}a_{11}\cos^{2}(\phi/2)
\label{Belltarget}
\end{equation}
For the $|BP_{1}>= a_{00}|00> + a_{01}|01> + a_{11}e^{i\phi}|11> $ state, the target function for the output $|<11|\Psi(t_{f})>|^{2}$ is:
\begin{eqnarray}
target_{BP_{1}} =  2|\frac{1}{3}-a_{01}^{2}|a_{00}^{2}a_{11}^{2} + 3|\frac{1}{3}-a_{00}^{2}|a_{01}^{2}a_{11}^{2} \\ \nonumber + 2a_{00}a_{11}\cos^{2}(\phi/2) 
\label{BP1target}  
\end{eqnarray} 
For the $|BP_{2}>= a_{00}|00> + a_{10}|10> + e^{i\phi}a_{11}|11>$ state, the target function for the output $|<11|\Psi(t_{f})>|^{2}$ is:
\begin{eqnarray} 
target_{BP_{2}} =  2|\frac{1}{3}-a_{10}^{2}|a_{00}^{2}a_{11}^{2} + 3|\frac{1}{3}-a_{00}^{2}|a_{10}^{2}a_{11}^{2} \\ \nonumber + 2a_{00}a_{11}\cos^{2}(\phi/2)  
\label{BP2target} 
\end{eqnarray}

Note that these target functions agree with the functions used for \underline{training}: that is, the training states $\frac{1}{\sqrt{2}}[ |00> + e^{i\phi}|11>]$, for $\phi: -\pi/2$ to $\pi/2$, had a target function given by $\cos^{2}(\phi/2)$; this is exactly what the training function in Equation \ref{Belltarget} reduces to, in the case $a_{00}=a_{11}=\frac{1}{\sqrt{2}}$. Similarly the target functions for the $|BP>$ states both reduce to the function tested on for the pure Bell states in the case of equal amplitudes of $\frac{1}{\sqrt{3}}$. Testing results for 550 randomly generated states of all three types, for all values of the angle $\phi$, are shown in Figure \ref{test3fig}. Agreement is quite good, even remarkable, considering that the system was trained only on 11 phase angles for an equal-amplitude Bell state. The average RMS error per pair over all 550 testing pairs after training is 0.0270.

With some confidence in our method, we now extend to the corresponding states of the two qubit system that also can have maximal entanglement, the EPR states, $|\Psi(0)>= a_{01}|01> + a_{10}e^{i \theta}|10>$. By symmetry, these states are ``the same'' as the Bell states; thus, we would expect that similar training ought to be able to map the phase shift to the projection onto the $|01>$ basis state. However, with \underline{no further training}, we were \underline{also} able to recover \underline{this} information! In other words, the neural net, trained to map $\phi$ information to the projection onto the basis state $|11>$, \underline{also} maps the $\theta$ information to the projection onto the $|01>$ basis state.   For equal amplitudes, $a_{01}=a_{10}=\frac{1}{\sqrt{2}}$, we again use the simple cosine function, $\cos^{2}(\theta/2)$; we use the analogous measure on the final state, $|<10|\Psi(t_{f})>|^{2}$. That is, the (input,output) pairs for equal amplitude EPR states are 
\begin{eqnarray}
 input =|\Psi(0)>= \frac{1}{\sqrt{2}}(|01> + e^{i \theta}|10> \\ \nonumber 
 output = |<10|\Psi(t_{f})>|^{2}\rightarrow target =\cos^{2}(\theta/2)
\end{eqnarray}

For non-equal amplitude EPR states, and for the analogous partially entangled EPR states, we employ exactly analogous target functions as with the Bell states. For the EPR state, $a_{01}|01> + a_{101}e^{i\theta}|10>$, we take the target function for the output $|<10|\Psi(t_{f})>|^{2}$ to be:  
\begin{equation}
target_{EPR} =  2(\frac{1}{2} - a_{01}^2)^{2} a_{10}^{2} + 2a_{01}a_{10}\cos^{2}(\theta/2)
\label{EPRtarget}
\end{equation}
For the $|EP_{1}>= a_{00}|00> + a_{01}|01> + a_{10}e^{i\theta}|10> $ state, the target function for the output $|<10|\Psi(t_{f})>|^{2}$ is:
\begin{eqnarray}
target_{EP_{1}} =  2|\frac{1}{3}-a_{00}^{2}|a_{01}^{2}a_{10}^{2} + 3|\frac{1}{3}-a_{01}^{2}|a_{00}^{2}a_{10}^{2} \\ \nonumber + 2a_{01}a_{10}\cos^{2}(\theta/2) 
\label{EP1target}  
\end{eqnarray} 
For the $|EP_{2}>= a_{01}|01> + a_{10}e^{i\theta}|10> + a_{11}|11>$ state, the target function for the output $|<10|\Psi(t_{f})>|^{2}$ is:
\begin{eqnarray} 
target_{EP_{2}} =  2|\frac{1}{3}-a_{11}^{2}|a_{01}^{2}a_{10}^{2} + 3|\frac{1}{3}-a_{01}^{2}|a_{11}^{2}a_{10}^{2} \\ \nonumber + 2a_{01}a_{10}\cos^{2}(\theta/2)  
\label{EP2target} 
\end{eqnarray}

\begin{figure}  
\centering
\includegraphics[width=2.5in]{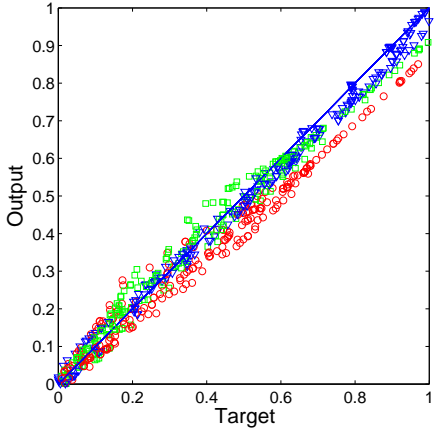}
\caption{Results for the testing set on the phase offset $\theta$, consisting of input states of three types: (1) $a_{01}|01> + a_{10}e^{i\phi}|10>$ (blue triangles); (2) $|EP_{1}(\theta)>= [a_{00}|00> + a_{01}|01> + e^{i\theta}a_{10}|10>$ (green squares), and (3) $|EP_{2}(\theta)>=a_{01}|01> + e^{i\theta}a_{10}|10> + a_{11}|11>$ (red circles.)  Again, we show the deviation of the output from the target function (given in the text.) Average RMS error per pair is 0.0489. The line shows the goal (perfect agreement.) \label{test4fig}}
\end{figure}

Results for testing on 550 randomly generated states of all three types are shown in Figure \ref{test4fig}. 

If we can recover phase offset information on both $|11>$ and $|10>$ projections, we ought to be able to do so on $|01>$. And so we can. Again we test only (\underline{no} additional training), using, this time, the projection onto the $|01>$ state, and looking for information about the phase offset term multiplied by that basis state. Our target functions are the exact analogues to the $|EPR>$ and $|EP_{1,2}>$ targets. For the EPR state with the $\xi$ offset, $a_{01}e^{i\xi}|01> + a_{10}|10>$, the target function for the output $|<01|\Psi(t_{f})>|^{2}$ is:  
\begin{equation}
target_{EPRx} =  2(\frac{1}{2} - a_{10}^2)^{2} a_{01}^{2} + 2a_{10}a_{01}\cos^{2}(\xi/2)
\label{EPRxtarget}
\end{equation}
For the $|EP_{3}>= a_{01}e^{i\xi}|01> + a_{10}|10> + a_{11}|11>$ state, the target function for the output $|<01|\Psi(t_{f})>|^{2}$ is:
\begin{eqnarray} 
target_{EP_{3}} =  2|\frac{1}{3}-a_{11}^{2}|a_{10}^{2}a_{01}^{2} + 3|\frac{1}{3}-a_{10}^{2}|a_{11}^{2}a_{01}^{2} \\ \nonumber + 2a_{10}a_{01}\cos^{2}(\xi/2)  
\label{EP4target} 
\end{eqnarray}
For the $|EP_{4}>= a_{00}|00> + a_{01}e^{i\xi}|01> + a_{10}|10> $ state, the target function for the output $|<01|\Psi(t_{f})>|^{2}$ is:
\begin{eqnarray}
target_{EP_{4}} =  2|\frac{1}{3}-a_{00}^{2}|a_{10}^{2}a_{01}^{2} + 3|\frac{1}{3}-a_{10}^{2}|a_{00}^{2}a_{01}^{2} \\ \nonumber + 2a_{10}a_{01}\cos^{2}(\xi/2) 
\label{EP3target}  
\end{eqnarray} 

\begin{figure}  
\centering
\includegraphics[width=2.5in]{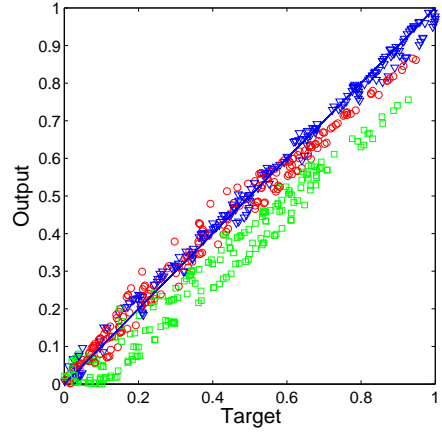}
\caption{Results for the testing set on the phase offset $\xi$, consisting of input states of three types: (1) $e^{i\xi}a_{01}|01> + a_{10}|10>$ (blue triangles); $|EP_{3}(\xi)>=a_{01}e^{i\xi}|01> + a_{10}|10> + a_{11}|11>$ (green squares), and (3) (2) $|EP_{4}(\xi)>= a_{00}|00> + e^{i\xi}a_{01}|01> + a_{10}|10>$   (red circles.)  Again, we show the deviation of the output from the target function. Average RMS error per pair is 0.0699. The line shows the goal (perfect agreement.) \label{test5fig}}
\end{figure}

Results are shown in Figure \ref{test5fig} for 550 randomly generated states. 

\section{Conclusion}

We have shown that a two-qubit quantum system, considered as a trainable quantum neural net, can compute its own phase offsets. The training is not difficult: the training set consisted of only 11 training pairs, of a single type, and the set was trained for only 10 epochs. Agreement is not perfect, and, doubtless, a more complicated function could be devised such that better agreement would be reached. But if we are considering inverting these functions, in order to perform the rotations that would enable our use of the entanglement estimator discussed in the Introduction, simplicity is also important. Because our method relies on the phase offset's being on a basis state which carries entanglement, it is not completely general; however, since our goal is to be able to estimate the entanglement of a general input state, it does not really matter, since no phase correction is necessary to an unentangled state, and would make no difference to the calculation if made.

Our previous work \cite{nabic, eb12}, which extended our work on entanglement in 2-qubit systems to n-qubit systems, seems to indicate that extension of our present results to multiple qubit systems should be possible without too much difficulty. The ease with which we are able to extract multiple angle information is encouraging. It should be not too difficult to perform the inverse rotations, and, thereby, to be able to form a good and reliable estimate for the entanglement with only a very few measurements, even for many-qubit systems. We are currently working on these calculations, and on the extension of our results to mixed systems.


\begin{thebibliography}{4}   

\bibitem{genentref}
M. Nielsen and I. Chuang, {\it Quantum computation and quantum information}. Cambridge: Cambridge University Press (2000).

\bibitem{eb08}
E.C. Behrman, J.E. Steck, P. Kumar, and K.A. Walsh, {\it Quantum algorithm design using dynamic learning}, Quantum Information and Computation {\bf 8}, pp. 12-29 (2008).

\bibitem{toth}
G. Toth and O. Guhne, {\it Detecting genuine multipartite entanglement with two local measurements}, Phys. Rev. Lett. {\bf 94}, 060501 (2005).

\bibitem{nabic}
E.C. Behrman and J.E. Steck, {\it Dynamic learning of pairwise and three-way entanglement}, in {\it Proceedings of the Third World Congress on Nature and Biologically Inspired Computing (NaBIC 2011)} (Salamanca, Spain, October 19-21, 2011. (Institute of Electrical and Electronics Engineers).

\bibitem{eb12}
E.C. Behrman and J.E. Steck, {\it Multiqubit entanglement of a general input state}, Quantum Information and Computation {\bf 13}, 36-53 (2013).

\bibitem{wootters}
 W.K. Wootters, {\it Entanglement of formation of an arbitrary state of two qubits}, Phys. Rev. Lett. {\bf 80}, pp. 2245-2248 (1998).

\bibitem{han}
 C-P Yang and S. Han, {\it Extracting an arbitrary relative phase from a multiqubit two-component entangled state}, 
Phys. Rev. A {\bf 72}, 014306 (2005).

\bibitem{yamamoto}
 T. Yamamoto, Yu.A. Pashkin, O. Astafiev, Y. Nakamura, and J.S. Tsai, 
{\it Demonstration of conditional gate operation using superconducting charge qubits},  Nature {\bf 425}, pp. 941-944 (2003). 

\bibitem{matlab}
MATLAB Simulink documentation notes [Online]. 
Available at http://www.mathworks.com/access/helpdesk/help/toolbox/simulink

\bibitem{lecun}
 Yann le Cun, {\it A theoretical framework for back-propagation} in 
{\it Proc. 1998 Connectionist Models Summer School,} D. Touretzky, G. Hinton, and T. Sejnowski, eds., Morgan Kaufmann, 
(San Mateo), pp. 21-28 (1988).

\bibitem{werbos}
 Paul Werbos, in {\it Handbook of Intelligent Control}, Van Nostrand Reinhold, p. 79 (1992).


\end{thebibliography}
\end{document}